\documentclass[aps,prl,twocolumn]{revtex4-1}
\usepackage{graphicx,color}  

\begin{document}

\title{(De)localization for Strong Disorder}
\author{H. Eleuch$^{1,2}$ and M. Hilke$^{1,3}$}
\address{$^1$Department of Physics, McGill University,  Montr\'eal, Canada H3A 2T8\\ $^2$Department of Physics, Universit\'{e} de Montr\'{e}al, Montr\'eal, Canada H3T 1J4\\
	$^3$Dahlem Center for Complex Quantum Systems, FU Berlin, Germany, 14195}
\email{heleuch@physics.mcgill.ca \& hilke@physics.mcgill.ca}


\begin{abstract}
In one dimension and for discrete uncorrelated random potentials, such as tight binding models, all states are localized for any disorder strength. This is in contrast to continuous random potentials, where we show here that regardless of the strength of the random potential, we have delocalization in the limit where the roughness length goes to zero. This result was obtained by deriving an expression for the localization length valid for all disorder strengths. We solved a non-linear wave equation, whose average over disorder yields the localization properties of the desired linear wave equation. Our results, not only explain the origin of the difficulty to observe localization in certain physical systems, but also show that maximum localization occurs when the roughness length is comparable to the wavelength, which is relevant to many experiments in a random medium.  	
\end{abstract}

\vspace{2pc}


\newcommand{\be}{\begin{equation}}
\newcommand{\ee}{\end{equation}}
\newcommand{\ba}{\begin{eqnarray}}
\newcommand{\ea}{\end{eqnarray}}
\newcommand{\heff}{H_{\rm eff}}
\newcommand{\ch}{{\cal H}}
\newcommand{\ce}{{\cal E}}

\maketitle

For more than half a century, thanks to the pioneering work by Anderson \cite{anderson1958absence}, we have taken for granted that strong disorder will lead to localization of all states, particularly in low dimensions and for uncorrelated disorder. Spatial localization is when an electron, an atom or even a photon, cannot propagate in a medium when disorder is large. Localization can also occur in time due to fluctuations \cite{fishman1982chaos} and was found to be relevant to the expansion of our universe \cite{brandenberger2012towards}. Formally, the propagation probability decays exponentially with the medium's length, which is known as Anderson localization. Even in the presence of interactions between particles, strong localization is expected to occur, which is now popularized with the term many-body localization \cite{basko2006metal}. Most of the theoretical work, has focused on discrete random equations, such as tight binding models, where the theoretical results for Anderson localization are confirmed by numerous numerical studies and the main message can be summarized as "all states are localized for uncorrelated disorder in one and two dimensions; in higher dimensions this is true for sufficiently strong disorder" \cite{abrahams1979scaling}. 

Experimentally, localization has been observed in many different systems, including electrons \cite{cutler1969observation}, photons \cite{schwartz2007transport} and atoms \cite{billy2008direct,roati2008anderson}. Moreover, applications are becoming increasingly important, particularly in random lasing \cite{wiersma2008physics,tureci2008strong} and optics \cite{skipetrov2014optical}. These different systems all share a similar underlying wave equation. Here,  instead of looking at discrete equations, we look directly at the continuous wave equation and show that for arbitrarily strong disorder, we have no localization when the roughness length vanishes, even in one dimension. While at intermediate roughness, localization is maximized. This is in stark contrast to discrete models, where no equivalent delocalization occurs. To obtain this result, we used a new approach based on solving an equivalent non-linear disordered wave equation. Our result explains why it is sometimes difficult to observe localization in certain physical systems when the roughness length is not of the same order as the wavelength.

Anderson localization ($AL$) has become an important phenomenon well beyond its original work on tight binding models with random potentials and couplings, which describe quantum particles or spins \cite{anderson1958absence}. $AL$ is important in photonic systems \cite{john1987strong,hilke2009seeing,skipetrov2014optical,topolancik2007experimental}, random lasers \cite{tureci2008strong,wiersma2008physics}, quantum information noise and entanglement \cite{crespi2013anderson}, atomic systems \cite{billy2008direct,roati2008anderson}, mechanical systems \cite{schwartz2007transport,weaver1990anderson}, biological systems \cite{epstein2006anderson}, cavity QED \cite{sapienza2010cavity}, as well as cosmology, where inflation is dependent on fluctuations \cite{brandenberger2012towards}. All these systems share a common underlying  wave equation, which can be written as
\be
[\partial_x^2+p^2(x)]\psi (x)=0,
\label{schroedinger}
\ee
where $\psi$ is the amplitude, $p(x)=\sqrt{\epsilon-v(x)}$, the classical momentum, $v(x)$ the random medium and $\epsilon$ the energy. We will restrict our attention here to the quasi-one dimensional situation, where the effect of disorder is the strongest. However, many of these results can be extended to higher dimensions and will be discussed elsewhere. Solutions to equation (\ref{schroedinger}) can be obtained for a random potential $v(x)$ that is not continuous. For instance, if $v(x)$ is written as a sum of delta functions or square wells, equation (\ref{schroedinger}) becomes equivalent to a tight binding equation studied by Anderson and others \cite{anderson1958absence,thouless1979ill,erdos1982theories,hilke1997delocalization}. The main result is the localization of all states if the potentials are uncorrelated, regardless of the strength of disorder. In the presence of correlations in the disorder, some states can be delocalized too \cite{flores1989transport,dunlap1990absence,flores1993absence,hilke1994local,Izrailev2012anomalous}. It is important to note here, that the minimum correlation length in tight binding models is limited to the smallest distance between impurities or orbitals. However, when several next nearest neighbors coupling elements are non-zero or when there is mixing between different energy bands, this induces effective correlations between neighboring onsite potentials. Hence, in this case too, a continuous potential model is more adapted.

\begin{figure}[h!]
	\centering
	\includegraphics[width=.5\textwidth]{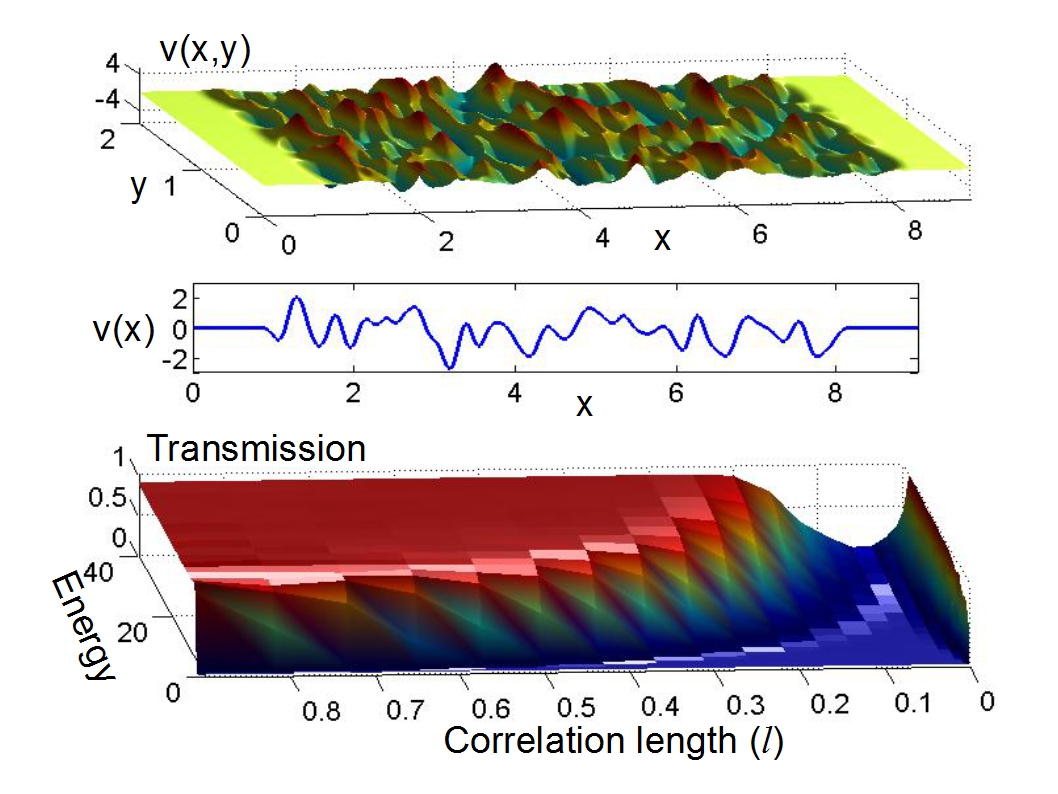}
	\caption{Top graph: a typical Gaussian disorder potential, $v(x,y)$ with correlation length $l$ connected by leads (in yellow). For a quasi 1D system in the lowest transverse mode the transverse potential can be integrated out to yield an effective 1D potential, $v(x)$ shown in the middle graph. The bottom graph shows the numerically calculated average transmission for such a random potential as a function of energy and correlation length $l$.}
	\label{conductance}
\end{figure}

\begin{figure}[h!]
	\centering
	\includegraphics[width=.5\textwidth]{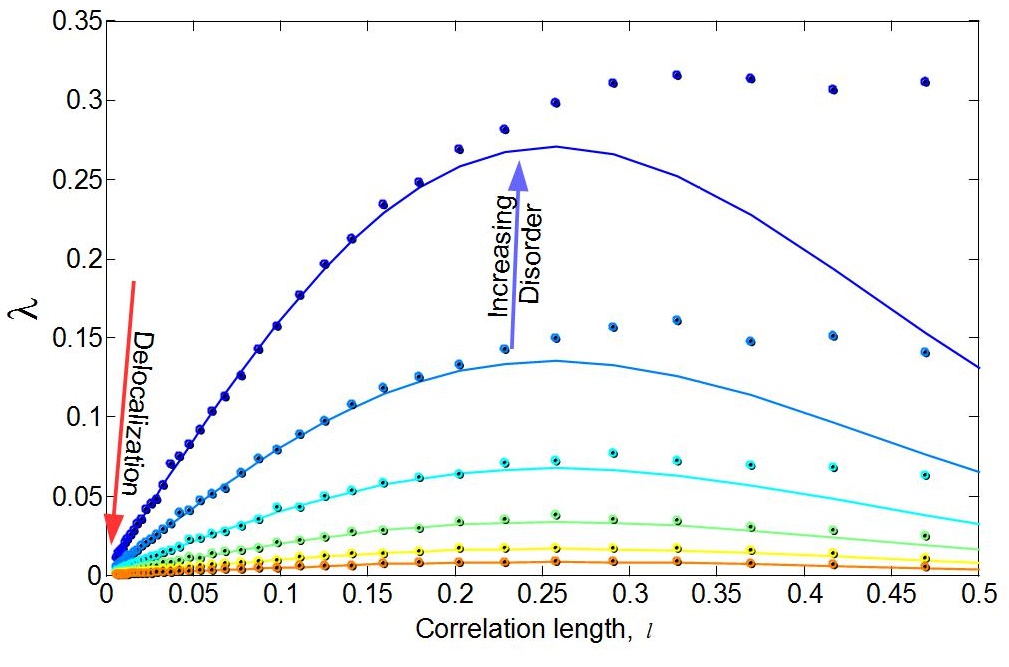}
	\caption{The Lyapounov exponent $\lambda$ as a function of correlation length $l$ for different values of the disorder strength (here $\sigma_v$ varies from 0.3 to 1.6 and $\epsilon=1.9$, hence the potential sometimes exceeds the energy). The binary correlator is taken to be Gaussian. The dots are the results obtained numerically for the decay of the transmission, with errors smaller than the size of the dots, while the lines are $2\lambda_w$ from expression (\ref{lambdaw}) with no fitting parameters. The factor 2 comes from the difference in defining $\lambda$ from the transmission versus the wavefunction amplitude.}
	\label{vanishing}
\end{figure}

When the potential $v$ is continuous, the situation changes. For instance, we can consider a typical quasi-one dimensional (Q1D) random potential of width $L_y$. Q1D means that at low energies, we can restrict ourselves to the one dimensional wave equation (\ref{schroedinger}), where only the lowest transverse mode is relevant and the Q1D solution is simply $\sin(\pi y/L_y)\psi(x)$, where $\psi$ is the solution for potential $v(x)=\int  \sin(\pi y/L_y)v(x,y)dy$ illustrated in figure \ref{conductance}. We can  consider the transport problem and evaluate numerically the transmission $T$, through such a potential assuming that we have perfect leads or wave guides at each end and represented in yellow in the figure. The numerical result is obtained by discretizing equation (\ref{schroedinger}) and then computing the disorder averaged transmission for a given system length. Care is taken in choosing a discretization parameter much smaller than both the disorder correlation length, $l$, and the wavelength. This leads to the non-monotonic behaviour of the transmission as a function of energy and correlation length $l$ shown in figure \ref{conductance}. At high enough energies and $l$, the transmission is maximum (1 in this model), while it is close to zero for a certain range of energies and $l$. This is the strong localization regime (AL), which is usually discussed in 1D random systems. In the opposite limit of vanishing correlation length $l$, the transmission is again maximum, which becomes a fully delocalized state at $l\rightarrow 0$. We show below, that this regime is robust with increasing disorder as represented in figure \ref{vanishing}.

\begin{figure}[h!]
	\centering
	\includegraphics[width=.5\textwidth]{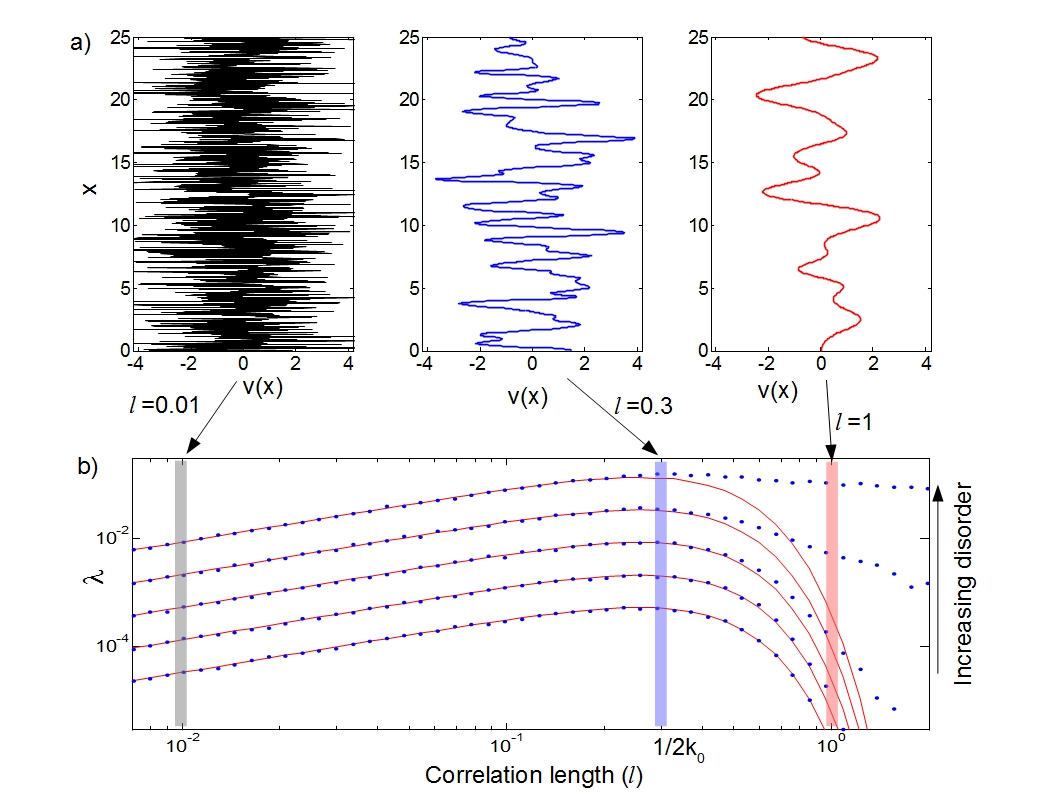}
	\caption{a) Examples of Gaussian disorder potentials with different correlation lengths ($l=0.01$, $l=0.3$, and $l=1$) but with the same standard deviation ($\sigma_v\simeq 1.6$). b) Correlation length ($l$) dependence of the Lyapounov exponent ($\lambda$) for different values of the disorder strength ($\sigma_v$ from 0.07 to 1.1) and $\epsilon=1.9$. The dots represent $\lambda$ obtained numerically from the transmission, while the red curves ($2\lambda_w$) are from the perturbative expression given in equation (\ref{lambdaw}).}
	\label{GaussianDis}
\end{figure}

For low disorder, we can understand the result in figure \ref{conductance} using the perturbative approach to disordered potentials \cite{Izrailev2012anomalous}. In this case it was found that when $\epsilon\gg v$, the inverse localization length, or Lyapounov exponent $\lambda_w$, is given by $\lambda_w\simeq\frac{\tilde{c}_v(2k_0)}{8\epsilon}$, where $c_v(x)=\langle v(0)v(x) \rangle$ is the binary correlator of Fourier transform $\tilde{c}_v(k)$, with $k_0=\sqrt{\epsilon}$ the wavenumber, and $\langle \cdot \rangle$ the disorder average. This result leads to a delocalization-localization-delocalization dependence as a function of the disorder correlation length $l$ shown in more detail in figure \ref{GaussianDis}. To compute $\lambda$, we considered a Gaussian correlated potential, i.e., $\langle v(0)v(x) \rangle=\sigma_v^2e^{-x^2/2l^2}$ with amplitude $\sigma_v$. Such a Gaussian binary correlator is obtained, for instance, when the potential, $v(x)$ is a sum of Gaussian impurities located at random sites. Representative potentials with different correlation lengths $l$ are shown in figure \ref{GaussianDis}a. The small disorder ($\sigma_v\ll\epsilon$) result for the Lyapounov exponent is given by
\be 
\lambda_w\simeq\frac{\tilde{c}_v(2k_0)}{8\epsilon}=\frac{l\sigma_v^2\sqrt{2\pi}}{8\epsilon}e^{-2k_0^2l^2},
\label{lambdaw}
\ee
which implies that for both $l\rightarrow 0$ and $l\rightarrow \infty$, $\lambda_w\rightarrow 0$, while maximum localization occurs for $l=1/2k_0$. In addition, we also have $\lambda_w\rightarrow 0$ for $\epsilon\rightarrow \infty$. This result differs substantially from the localization behaviour of the 1D disordered Anderson tight binding model (AM), where localization occurs for all energies \cite{thouless1979ill}. Moreover, in the AM, localization ($\lambda$) does not vanish at the smallest correlation length. However, long range correlations in the $AM$ model can also lead to delocalization \cite{de1998delocalization,shima2004localization}, similarly to the continuous case shown here. In general, there is a decrease of the Lyapounov exponent with energy as seen by the $\epsilon^{-1}$ prefactor in equation (\ref{lambdaw}), which is also true in higher dimensions \cite{filoche2012universal}. However, correlations such as the roughness of the potential can override this behavior due to the exponential dependence on $l$, which in some cases can even lead to delocalization at small energies as illustrated in figure \ref{conductance}. 

\begin{figure}[h!]
	\centering
	\includegraphics[width=.5\textwidth]{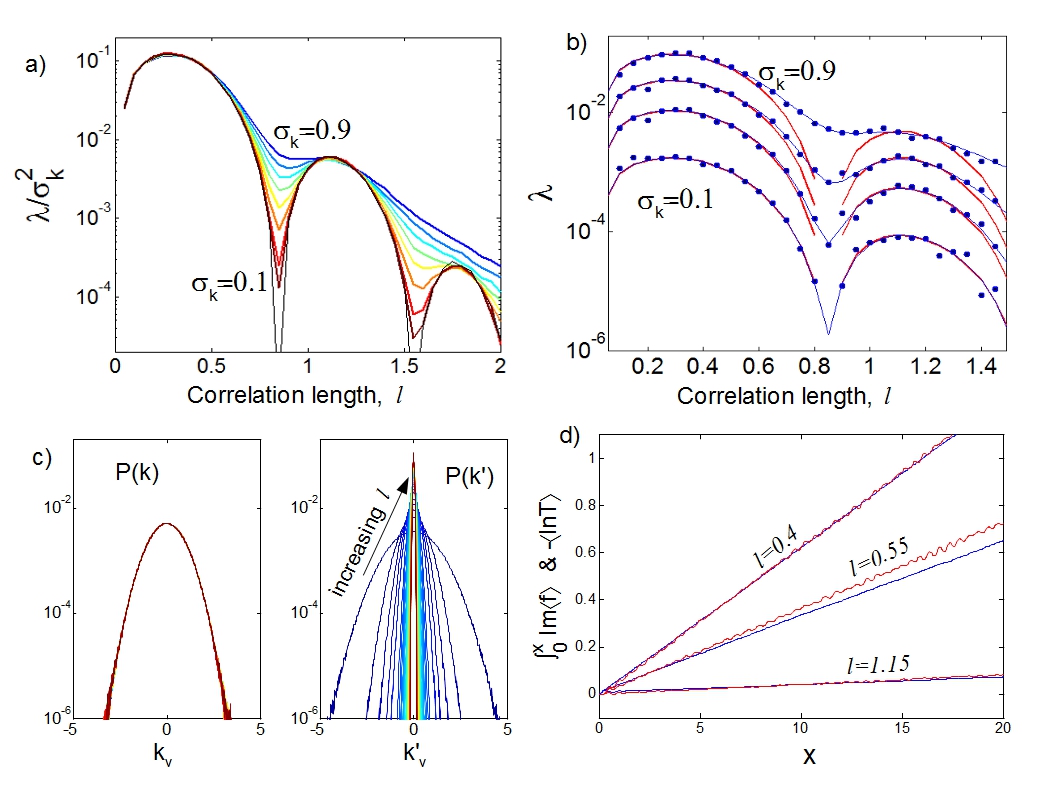}
	\caption{Graph a) represents the Lyapounov exponent normalized by the disorder strength as a function of $l$. The colored lines are the numerical results, while the black line is $2\lambda_w=\tilde{c}_v(2k_0)/4\epsilon$ from the perturbative expression (at the resonances $\lambda_w=0$). In b) the dots are from  expression (\ref{main}) evaluated numerically, while the blue lines are the numerical transmission results and the red lines are $2\lambda_w$ from the perturbative expression. c) The distributions of the potential $k_v(x)$ (left) and $k_v'(x)$ for $l$ ranging form 0.05 to 2. d) Plot of the increase of $\int_0^x \langle f_v\rangle $ with $x$ (equation \ref{f}) shown in red, and the blue lines correspond to minus the logarithm of the transmission $-\langle \ln T(x)\rangle$ evaluated numerically. A more detailed disorder strength dependence around the resonance is discussed in the appendix. In all figures, we used $\epsilon=40$.}
	\label{increasingdisorder2}
\end{figure}

At large disorder and when the correlation length is large, expression (\ref{lambdaw}) breaks down as seen in figures \ref{vanishing} and \ref{GaussianDis}. To understand the localization behaviour in this regime, which is relevant to many experiments, we need to go beyond the perturbative result, which brings us to our new approach to localization physics. The main idea is to solve an analogue to equation (\ref{schroedinger}) but with an additional non-linear term:
\be
[\partial_x^2+p^2(x)]\psi (x)=-\left[
\psi(x)^{-1}(-i\partial_x-p(x))\psi(x)\right] ^{2}\psi (x).
\label{ERS}
\ee
Interestingly, there exists an exact solution to equation (\ref{ERS}), which can be expressed in terms of the integral solution $\psi(x)= \psi(0)e^{i\int_0^{x}f(x')dx'}$ \cite{eleuch2010new}, where
\be
f(x)=p(x)-\underbrace{e^{-2iP_I(x)}\int_0^xk_v'(x')e^{2iP_I(x')}dx'}_{f_v(x)}.
\label{f}
\ee
Here $P_I(x)=\int_0^x p(x')dx'$ is the integrated classical momentum. If the right side term in equation (\ref{ERS}) vanishes, we recover our original equation (\ref{schroedinger}). Since this term is the difference between the classical and quantum momentum, we expect this term to be small and to vanish with disorder averaging (see appendix). Hence the localization behavior of the non-linear equation (\ref{ERS}) will describe the localization behavior of the linear equation (\ref{schroedinger}). The last term $f_v$ in equation (\ref{f}) describes the memory effect of the wave propagation, expressed as an integral. For clarity, we have expressed the disorder dependence of $p(x)$ in terms of $k_v(x)=p(x)-k_0$ with average 0 and variance $\sigma_k^2=\langle k_v^2\rangle$. $k_v'(x)$ is the spatial derivative of $k_v(x)$, which we assume to be finite and which scales as $1/l$. We only consider the case where $k_v'(x)$ remains finite, hence no discontinuous potentials. From here on, all the results will be expressed in terms of $k_v(x)$ rather than $v(x)$. The reason is that at high disorder this is the relevant quantity, while at low disorder they are proportional, since $v(x)\simeq -2k_v(x)\sqrt{\epsilon}$. The term $f_v(x)$ contains the physics relevant to the localization behaviour. Its average over all disorder configurations can be expressed in terms of a new correlation function $c_p(x)$:
\be
\langle f_v(x) \rangle=\int_0^xdx'e^{-2ik_0(x-x')}\underbrace{\langle k'_v(x') e^{-2i\int_{x'}^xk_v(x'')dx''}\rangle}_{c_p(x,x')=c_p(x-x')}.
\label{cp}
\ee
The real part of the integral $\int^x f(x')dx'$, which appears in the wavefunction solution, determines the wavenumber, while the imaginary part corresponds to the exponential dependence of the wavefunction. Hence we expect the disorder average of the imaginary part to be related to localization. Indeed, figure \ref{increasingdisorder2} shows the linear increase with x of $\int^x_0 \Im\langle f_v(x')\rangle dx'$ with proportionality coefficient $\lambda$. More precisely, we have
\be
\lambda=\Im\frac{1}{X}\int_0^Xdx\langle f_v(x)\rangle\bigg\vert_{X\rightarrow\infty} =\Im \int_0^\infty dye^{-2ik_0y}c_p(y),
\label{main}
\ee
assuming $k_0$ real. This follows from the exponential dependence of the wavefunction, which determines $\lambda$ and can be expressed as $\lambda=\frac{1}{X}\ln\langle\vert\psi(X)/\psi(0)\vert\rangle_{X\rightarrow\infty}$. Equation (\ref{main}) is the main analytical result of this paper and is valid for all disorder strengths and correlations. It's validity is illustrated in figure \ref{increasingdisorder2}. For $c_p(x)$ symmetric we have $\lambda=\Im\tilde{c}_p(2k_0)/2$. 

For arbitrary potentials, the correlator $c_p(x)$ can be quite difficult to evaluate. However, it is possible to describe the localization behaviour in different important limits. The limit, where $l$ is large and where the perturbative expression breaks down at large disorder, is important to understand for many experimental systems, like in semiconductors in the presence of long range strong Coulomb potentials \cite{timm2002correlated} or for scattering in photonic crystals \cite{o2010loss}. In this limit, where we consider a disorder potential characterized by a large but finite correlation length $l$ and assuming that $\vert x\vert\ll l$ we can write (see appendix)
\be
c_p(x)
\simeq \langle  e^{-2ixk_v(0)}k'_v(0)e^{-ix^2k'_v(0)}\rangle\simeq
i\tilde{P}(2x)\tilde{P}'_p(x^2).
\label{xsmall}
\ee
We used that $k'_v(0)$ is not correlated to any function of $k_v(0)$, since $\langle k'_v(0)k_v^n(0)\rangle=0$ for any positive integer $n$, when assuming random Gaussian impurities. The term $\tilde{P}'_p(x^2)\simeq\langle [(k_v(x)-k_v(0))/x] e^{-ix(k_v(x)-k_v(0))}\rangle=\sum_{n=0}-i(-ix)^{n-1}\langle [k_v(x)-k_v(0)]^{n+1}\rangle$ is related to the moment generating two point correlator and the Fourier transform of the distribution function of $k'_v$ which sharpens for larger $l$ as shown in figure \ref{increasingdisorder2}c. It can be written as $\tilde{P}'_p(x^2)\simeq 2[c_k(x)-c_k(0)+hm(x)]$, where $hm(x)$ are higher moments. In terms of notation, $\tilde{P}'(k)\equiv\partial_k\tilde{P}(k)$. $\tilde{P}(2x)=\langle e^{-2ixk_v(0)}\rangle$ is the Fourier transform of the distribution of $k_v$, which is largely independent of $l$ as seen in figure \ref{increasingdisorder2}c. The binary correlator, $c_k(x)=\langle k_v(0)k_v(x) \rangle$ and consequently $c_p(x)\simeq 2i[c_k(x)-c_k(0)]$ in the lowest order of the disorder strength. This result can also be obtained directly from the Taylor expansion of the exponential term in $c_p(x)$ and keeping only the first non-zero term.

Assuming that $k_0l\gg1$, we can evaluate the Fourier transform of $c_p(x)$ at $2k_0$, which gives rise to the following convolution: 
\be
\tilde{c}_p(2k_0)
= 2i\int dk P(k)[\tilde{c}_k^*(2k_0-2k)+\widetilde{hm^*}(2k_0-2k)].
\label{llarge}
\ee
The first term is the distribution function of the disorder potential of width $\sigma_k=\sqrt{\langle k_v^2\rangle}$ centred at zero and nicely illustrates what happens with increasing disorder. For small enough disorder,  $P(k)$ is simply a delta function centred at $k=0$, hence using equation (\ref{main}) and dropping the higher moments, we have $\tilde{c}_p(2k_0)\simeq 2i \tilde{c_k}(2k_0)$ and $\lambda=\Im\tilde{c}_p(2k_0)/2\simeq\tilde{c}_k(2k_0)\simeq\tilde{c}_v(2k_0)/4\epsilon=2\lambda_w$, which is twice the value of the perturbative result obtained in ref. \cite{Izrailev2012anomalous}. This factor of 2 is due to the different averaging method. Indeed, in one dimensional disordered systems, we have for $X\rightarrow\infty$, $\lambda=\ln\langle \vert\psi(X)\vert\rangle/X=2\langle\ln\vert\psi(X)\vert\rangle /X=2\lambda_w$, because $\vert\psi(X)\vert$ follows a log-normal distribution \cite{pendry1994symmetry}. Hence our result is equivalent to averaging the wavefunction directly. For larger disorder the Lyapounov exponent becomes the convolution of the low disorder value at $2k_0$ averaged around $2k_0\pm 2\sigma_k$. 

This is best illustrated in the context of a disorder potential, where the Fourier transform of the binary correlator, has resonances. Such a potential can be obtained, for example, by starting with an uncorrelated potential and then making it smooth over a length scale $l$ (see details in the appendix and ref. \cite{cleveland1979robust}). This leads to strong minima or resonances in the Fourier transform amplitude as reflected in figure \ref{increasingdisorder2}. Precisely at the resonance ($l=0.85$ in figure 4), the perturbative approach gives $\lambda_w=0$. With increasing disorder, the resonances, where $\lambda$ is minimum, first broaden (described by a convolution) then reach a high disorder regime. At high disorder the full correlation function $\tilde{c}_p(2k_0)$ needs to be evaluated, which involves all moments. The second moment (or binary correlator) $\tilde{c}_k(2k_0)\sim\vert\tilde k_v(2k_0)\vert^2$ is simple to compute, since it is proportional to the squared absolute value of the Fourier transform of the disorder potential.  

In the other limit, where the correlation length $l$ is small, we can see in figures \ref{GaussianDis} and \ref{increasingdisorder2}, that the behavior is largely independent on the disorder strength (except for the multiplicative factor). We can understand this result by first looking at the behavior of $c_p(x)$ in the limit $\vert x\vert\gg l$ , where
\be
c_p(x)
\simeq\langle k'_v e^{-i l^2k'_v}\rangle\langle e^{-2i lk_v}\rangle^{\vert x\vert/l}\simeq i\tilde{P}'_p(l^2)\tilde{P}(2 l)^{\vert x\vert/l}.
\label{xlarge}
\ee
For $l\rightarrow 0$ the exponential decay of $c_p(x)$ vanishes and $c_p(x)\simeq i\tilde{P}'_p(l^2\rightarrow 0)$. For most disorder distributions we can write $\tilde{P}'_p(l^2\rightarrow 0)\sim -\sigma_k^2$ (for a Gaussian distribution the proportionality coefficient is one and $\tilde{P}(2 l)^{\vert x\vert/l}\rightarrow 1$ for $l\rightarrow 0$). Hence, to determine the Fourier transform of $c_p(x)$ when $l\rightarrow 0$, we can consider $c_p(x)\sim -i\sigma_k^2$ for $\vert x\vert>l$ and $c_p(x)\sim -i\sigma_k^2x^2/l^2$ for $\vert x\vert<l$ using equation (\ref{llarge}). This leads to $\tilde{c}_p(2k_0)\simeq il\sigma_k^2$ for  $l\rightarrow 0$ ($\lambda\sim l\sigma_k^2$) and the result is valid for any disorder strength and only the proportionality coefficient will depend on the disorder distribution. Therefore, for arbitrary disorder strength, $\lambda$ will vanish linearly with vanishing $l$. This delocalization can be understood, as the zero average of the disorder potential within a wavelength. On the other hand, localization is the strongest when the wavelength is comparable to $l$ and then $\lambda$ decays again at large $l$. For very high disorder ($\sigma\gg \epsilon$), we find numerically that $\lambda\sim \sigma^2l$ still applies for vanishing $l$ but then remains constant for larger $l\gg k_0^{-1}$.

To conclude, the (de)localization behavior has important implications to our understanding of low dimensional systems. Often it is assumed that in one dimension, all randomness will localize, but as we have shown here this is only the case for $k_0l\simeq 1$. For instance, in widely studied systems, such as GaAs bases heterostructures, the disorder correlation length can be of the order of $l\simeq 100nm$, while the Fermi wave length is only about $k_0^{-1}\simeq 10nm$ \cite{renard2004large}. Hence we expect localization effects to be strongly suppressed when dominated by long range disorder. Depending on the disorder correlation, this suppression can be exponential (for a Gaussian binary correlator) or quadratic for an exponential correlator. In atomic systems this effect is important too, since usually $l>100nm$ and the atomic de Broglie wavelength can be very small \cite{billy2008direct}. In the other extreme, of very short range disorder, like alloy scattering, the disorder correlation length is of the order of $l\simeq 0.1nm$ (the atomic distance). For a typical Fermi wavelength of 10nm, this leads to an increase of the localization length by two orders of magnitude. A similar situation arises in photonic systems, where the wavelength is of the order of $500nm$, but if the disorder correlation length is much smaller, then no localization can be observed. We believe that the suppression of localization $\lambda\sim l\sigma^2$ at small $l$ is not necessarily unique to the continuous potentials we considered here, but is likely to occur in other systems too. For instance, the equivalent tight binding model, with discretization $a$, would renormalize the disorder potential by $a^2v$ for unit bandwith, which suggests delocalization for small $a$. More generally, any potential with fixed $\sigma_v$ but vanishing integral over the wavelength is`likely to lead to delocalization.

Summarizing, we have shown that the localization behaviour of the standard disordered wave equation can be computed for all disorder strengths and correlation lengths using the disorder average of an approximate non-linear wave equation. This has important implications on our understanding of disordered systems and its applications \cite{karbasi2014image} as well as cosmological fluctuations \cite{harrison1970fluctuations,brandenberger2012towards}.

{\bf Acknowledgements:} The authors acknowledge helpful discussions with Nykolay Makarov and Arkadii Krokhin as well as financial support from FQRNT and INTRIQ.

H.E. and M.H. contributed equally to this work.

\bibliography{AL}


\section{Appendix}

\subsection{Derivation of equation (4)}

Equation (3) reads
\be
[\partial_x^2+p^2(x)]\psi (x)=-\left[
\psi(x)^{-1}(-i\partial_x-p(x))\psi(x)\right] ^{2}\psi (x).
\ee
taking $\psi=e^{i\int^xf(x')dx'}$ leads to 
\be
i\partial_xf(x)+2p(x)^2-2p(x)f(x)=0
\ee
Assuming $f(x)=p(x)+\eta(x)$ we have the following first order differential equation for $\eta$,
\be
\partial_x[\eta(x)+p(x)]+2i\eta(x)p(x)=0, 
\ee
with solution for $\eta(0)=0$ and $P_I(x)=\int^x_0p(x')dx'$,
\be
\eta(x)=-e^{-2iP_I(x)}\int^x_0e^{2iP_I(x')}(\partial_{x'}p(x'))dx'.
\ee
Defining $p(x)=k_v(x)+k_0$ we obtain equation (4). It is important to note that no approximations have been made beyond considering equation (3) instead of equation (1).

The right hand side term in equation (3) is shown in figure 5 to be small and to vanish after disorder averaging.

\subsection{Derivation of equation (7)}
We have from equation (5)
\be
c_{p}(x)=\langle k'_{v}(0)e^{-2i\int_{0}^{x}k_{v}(x^{\prime })dx^{\prime
}}\rangle.
\ee
Expanding $k_v(x)\simeq k_v(0)+xk'_v(0)$ for $\vert x\vert\ll l$, yields
\ba
c_p(x)
&\simeq& \langle  e^{-2ixk_v(0)}k'_v(0)e^{-ix^2k'_v(0)}\rangle\nonumber\\
&\simeq&
\langle  e^{-2ixk_v(0)}\rangle\langle k'_v(0)e^{-ix^2k'_v(0)}\rangle\
\ea
since $k'_v(0)$ is not correlated to any function of $k_v(0)$ because $\langle k'_v(0)k_v^n(0)\rangle=0$ for any positive integer $n$, when assuming random Gaussian impurities. The last term can now be expressed in terms of the distribution function of the disorder potential $k_v$, i.e., $\tilde{P}(2x)=\int dk_vP(k_v)e^{-2ixk_v}=\langle  e^{-2ixk_v}\rangle $. Similarly we have $i\tilde{P_p'}(x^2)=\int dk'_vP_p(k'_v)k'_ve^{-ix^2k'_v}$, where $P_p$ is the distribution function of the disorder potential $k'_v$. Hence we obtain equation (7). 

\vspace{.2cm}

\subsection{Disorder strength dependence around the resonance}

We used two different techniques to obtain a random potential characterized by a correlation length $l$. In figures 1-3 we used the sum of Gaussian impurities with random amplitudes and located at random sites. The number of impurities scales as $1/l$ in order for the standard deviation of the potential to be independent of $l$. The advantage of this potential is that computing the binary correlator is very simple as seen in equation (2). In figures 4 and 5 we used a random potential obtained by smoothing an uncorrelated potential over $l$ neighbors using the local regression smoothing process with tri-cube weight functions [35]. This produces a smooth potential with a characteristic correlation length $l$. Here we have no simple expression for the binary correlator or its Fourier transform, which has to be computed numerically. However, the Fourier transform of this correlator has resonances where the Fourier transform vanishes, which corresponds to the resonances seen in figure 4. A typical realization is shown in figure 5 as well as the dependence of $\lambda$ on disorder strength.
 
\begin{figure}[h!]
	\centering
	\includegraphics[width=.5\textwidth]{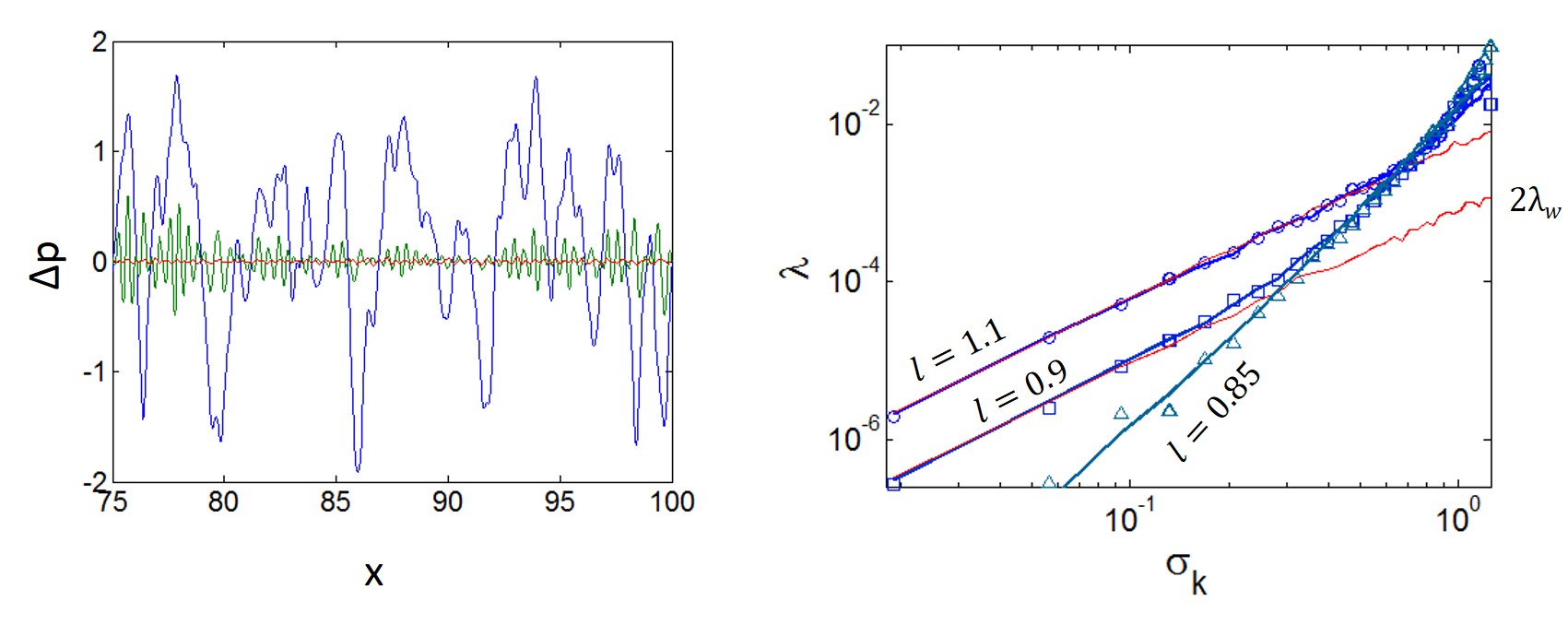}
	\caption{Left: the blue line shows the spatial dependence of $k_v(x)$ for a particular disorder realization. The green line is the difference between the quantum and the classical momentum ($\psi^{-1}(x)[-i\partial_x-p]\psi(x)$), while the red line is the disorder averaged difference (1000 realizations) $\langle\psi^{-1}(x)[-i\partial_x-p]\psi(x)\rangle$. Here we used $k_0=6.3$, $l=1.15$ and $\sigma_k=0.9$. Right: The Lyapounov exponent as a function of the disorder strength $\sigma_k$ around a resonance ($l=0.85$). The blueish lines are the numerical $\lambda$ from the transmission, the blueish symblols are from our approach (equation (6)), while the red lines are from the perturbative approach ($2\lambda_w$). For $l=0.85$ the perturbative approach gives $\lambda_w=0$ regardless of disorder.}
	\label{highdis}
\end{figure}


\end{document}